\documentclass[10pt,letterpaper,twocolumn]{article} 

\usepackage{ol2}
\usepackage[draft]{hyperref}
\usepackage{amsmath}
\usepackage{amssymb}

\begin{document}

\twocolumn[ 

\title{Surface-enhanced Raman spectroscopy (SERS): nonlocal limitations}

\author{G. Toscano,$^1$ S. Raza,$^{1,2}$ S. Xiao,$^1$ M. Wubs,$^1$ A.-P. Jauho,$^3$ S.I. Bozhevolnyi,$^4$ and N.A. Mortensen$^1$}

\address{$^1$Department of Photonics Engineering, Technical University of Denmark, DK-2800 Kgs. Lyngby, Denmark\\
$^2$Center for Electron Nanoscopy, Technical University of Denmark, DK-2800 Kgs. Lyngby, Denmark\\
$^3$Department of Micro and Nanotechnology, Technical University of Denmark, DK-2800 Kgs. Lyngby, Denmark\\
$^4$Institute of Sensors, Signals and Electrotechnics, University of Southern Denmark, DK-5230 Odense, Denmark}


\begin{abstract}
Giant field enhancement and field singularities are a natural consequence of the commonly employed
local-response framework. We show that a more general nonlocal treatment of the plasmonic response
leads to new and possibly fundamental limitations on field enhancement with important consequences for
our understanding of SERS. The intrinsic length scale of the electron gas serves to smear out
assumed field singularities, leaving the SERS enhancement factor finite even for geometries with
infinitely sharp features. For silver nano-groove structures, mimicked by periodic arrays of
half-cylinders (up to 120\,nm in radius), we find no enhancement factors exceeding ten orders of magnitude ($10^{10}$).
\end{abstract}
] 

\begin{figure}[b!]
\centering \includegraphics[width=0.6\columnwidth]{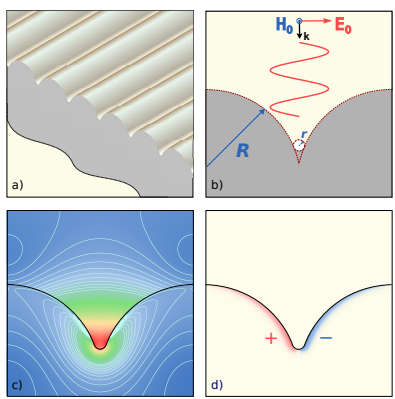}
\caption{(a) Groove structure formed
by an infinite periodic array of half-cylindrical nanorods. (b) Cross
section of the unit cell. (c) and (d) Typical electric-field intensity and charge distributions for a dipole mode.}
\label{plot1}
\end{figure}

While the Raman response of (bio-)molecules is inherently weak, nanostructures may be used to
tailor and tremendously enhance the light-matter interactions. This is the key electromagnetic
element of surface-enhanced Raman spectroscopy (SERS)~\cite{Moskovits:1985}. In particular, metallic
nanostructures~\cite{Lal:2007} are known to support plasmonic field-enhancement phenomena which are
beneficial for SERS~\cite{Kneipp:2007}. In many cases, field singularities arise in geometries with
abrupt changes in the surface topography. While such singularities constitute the basic
electromagnetic mechanism behind SERS, the singularities are on the other hand an inherent
consequence of the common local-response approximation (LRA) of the plasmons~\cite{Luo:2010}. In this Letter,
we relax this approximation and allow for nonlocal dynamics of the plasmons. To illustrate the
consequences we revisit the model geometry in~\ref{plot1}, initially put forward by
Garc\'ia-Vidal and Pendry~\cite{Garcia-Vidal:1996} to qualitatively explain the electromagnetic
origin of the large enhancement factors observed experimentally. The metallic surface topography is
composed of a periodic structure of infinitely long metallic half-cylinders of radius~$R$, resting
shoulder-by-shoulder on a semi-infinite metal film. The steep trenches or grooves support
localized-surface plasmon resonances (LSPR). Near the bottom of the groove the surfaces of the two
touching half-cylinders become tangential to each other and a field singularity forms within the
traditional LRA of the dielectric function. In the common treatment, the field enhancement thus
eventually turns infinite~\cite{Romero:2006} while it remains finite, albeit large, in any experiment
reported so far. Geometrical smoothening is known to remove the singularity within the LRA and in
quantitative numerical studies a rounding needs to be added to make numerical convergence
feasible~\cite{Moreno:2006,Xiao:2008}. Thus, within the LRA framework the field enhancement would
just grow without bound the sharper one could make the geometry confining the plasmon oscillations.
Nonlocal effects have been shown to result in large blueshifts and considerably reduced field
enhancements (as compared to a local description) in conical tips~\cite{Ruppin:2005}, metallic dimers involving small gaps below a
few nanometers~\cite{Garcia-de-Abajo:2008,Toscano:2012}, or even vanishing gaps~\cite{Fernandez-Dominguez:2012}. \emph{What is the
limit in field enhancements that can be achieved with (geometrically) ideal structures?} This
question is important not only from the fundamental but also from applied perspective, as the answer
to it would allow one to determine technological tolerances in fabrication of nanostructures
designed for achieving record-high field enhancements. In this Letter we show how nonlocal response
introduces a new intrinsic length scale that serves to remove the field singularities, leaving field
enhancements finite even in geometries with arbitrarily sharp changes in the surface topography. For
the particular geometry of Fig.~\ref{plot1} we evaluate $\gamma(\mathbf{r}, \omega)=   \left|
\mathbf{E}(\mathbf{r},\omega)   \right|^4/\left|\mathbf{E}_\textrm{0}(\omega)\right|^4$ and find no (surface averaged) SERS enhancement factors
$\big<\gamma\big>$ exceeding ten orders of magnitude.

\begin{figure}[t!]
\centering \includegraphics[width=0.7\columnwidth]{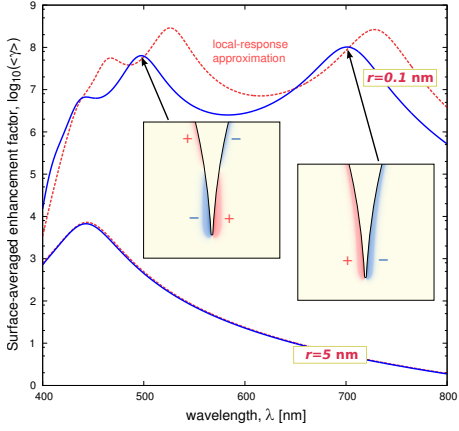}
\caption{Surface-averaged SERS enhancement factor $\left<\gamma\right>$ for the case of
$R=75$~nm with $r=0.1$\,nm (upper curves) and $r=5$\,nm (lower curves). For
comparison, the dashed lines show the results of the commonly employed local-response approximation.}\label{plot2}
\end{figure}

The electromagnetic response of a metal is commonly divided into intraband contributions~\cite{Rodrigo:2008} and the dispersive Drude free-electron
response $\varepsilon_D(\omega)=1+i\frac{\sigma}{\varepsilon_0 \omega}=1-\frac{\omega_p^2}{\omega(\omega+i/\tau_D)}$, where $\sigma$ is the complex conductivity also appearing in Ohm's law ${\mathbf J}=\sigma{\mathbf E}$. We relax the latter local-response constitutive equation and turn to a linearized hydrodynamic nonlocal treatment~\cite{Pitarke:2007,Garcia-de-Abajo:2008,Raza:2011,Toscano:2012} where the usual Maxwell wave equation is coupled to a hydrodynamic equation for the current density, see Ref.~\cite{Toscano:2012} for the full details of our numerical approach. This is the simplest non-trivial extension of the common LRA Drude model, which in addition to the
usual metal parameters ($\omega_p$, $\tau_D$, etc.) now also carries information about the
kinetics of the charge carriers at the Fermi level. The strength of the
nonlocal correction to Ohm's law depends on the Fermi velocity $v_F$ which introduces a new length
scale, being a factor $v_F/c$ of the free-space wavelength $\lambda=2\pi c/\omega$. For the noble
metals, $v_F/c$ is of the order $10^{-2}$ which explains the overall success of the LRA. However,
when exploiting plasmonics at the true nanoscale, effects due to the nonlocal dynamics start to
manifest themselves. Field-enhancement structures turn out to be prime examples of this.

\begin{figure}[b!]
\centering \includegraphics[width=0.7\columnwidth]{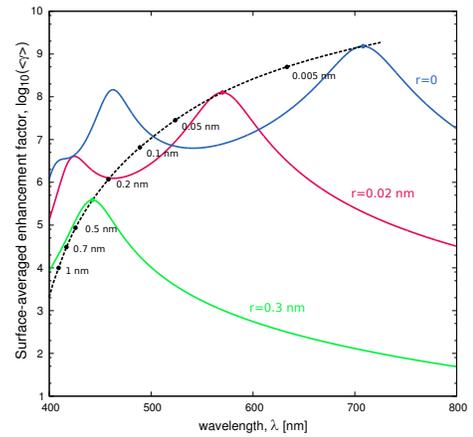}
\caption{Surface-averaged SERS enhancement factor $\left<\gamma\right>$ for the case of $R=15$~nm
and with $r$ varying from $1$\,nm to $0$\,nm. The dashed line connecting fundamental dipole
resonances for different values of $r$ serves as a guide to the eyes, clearly illustrating both a
redshift and the saturation effect in the field enhancement as $r\rightarrow 0$.}\label{plot3}
\end{figure}

We consider the metallic groove structure shown in Fig.~\ref{plot1} which has previously been
considered as a model system to mimic corrugated metal surfaces~\cite{Garcia-Vidal:1996}.
Alternatively, it may be viewed as a model for arrays of the more recent groove or channel
waveguides~\cite{Moreno:2006,Sondergaard:2010}. In our numerical study, the structure is excited by
an incoming plane wave $\mathbf{E}_\textrm{0}(\omega)$, normal to the substrate and with the field
polarized perpendicularly to the axis of the half-cylinders, i.e. across the groove cross section.
Noble metals are common choices for plasmonics and in the following we focus our attention on
silver~\cite{Rodrigo:2008}. The grooves have been shown to support LSPRs~\cite{Moreno:2006} which we have previously
explored in the context of SERS, using a LRA and with the necessary addition of geometrical
smoothening~\cite{Xiao:2008}. To quantify the SERS effect and the consequences of nanoscale spatial
dispersion, we solve the nonlocal wave equation numerically~\cite{Toscano:2012}. As an example of
our results, Fig.~\ref{plot2} shows the spectral dependence of $\left< \gamma \right>$ throughout
the visible regime for groove structures with $R=75$\,nm  and with a radius of curvature of the
crevice given by $r=0.1$\,nm. The LSPR at $\lambda=700$~nm allows the (surface-averaged) Raman rate
to be enhanced by a factor of $10^{8}$. For comparison, the dashed line shows results when treating
the plasmonic response within the common LRA. In both cases, the resonant behavior is well
pronounced, being caused by interference of the incoming field with the gap surface plasmon mode
reflected at the bottom, similarly to that described for V-grooves~\cite{Sondergaard:2010}. As a
general fingerprint of nonlocal response, the peak is blueshifted compared to the expectations from
a local-response treatment of the problem (this happens due to a decrease in the gap plasmon index
caused by nonlocal effects~\cite{Garcia-de-Abajo:2008}). In this particular case, the LSPR by the
common treatment is off by more than 25\,nm which illustrates the importance of nonlocal effects for
quantitative SERS predictions. Even more importantly, the common LRA is seen to significantly
overestimate the enhancement factor; for some wavelengths by more than one order of magnitude. The
large quantitative differences between the nonlocal treatment and the traditional LRA are associated
with changes in the induced-charge distribution (insets of Fig.~\ref{plot2}). In the common
treatment, the charge is strictly a surface charge while in the general nonlocal case the intrinsic
scale $v_F/\omega$ serves to spatially smear out the charge distribution. Effectively, this smearing
increases the electric field penetration into metal (silver) and thereby increases the field
absorption (ohmic loss) and damping of resonant oscillations. Interpreting the field enhancement in
a capacitor picture, the finite thickness of the charge distribution near the surface increases the
effective separation (beyond that given by the metal-surface geometry) and consequently the
capacitor supports a lower electrical field compared to in the LRA. In general, the intrinsic length
scale of the electron gas allows to resolve the field also in the proximity of very sharp
corners and tips. On the other hand, by relaxing the sharpness of the trench the influence of
spatial dispersion becomes less pronounced, as illustrated in Fig.~\ref{plot2} in the lower set of
curves ($r=5$\,nm) where the LRA accounts well for the results obtained from a full nonlocal
treatment. We note a drastic change in the field enhancement spectrum, with the fundamental
resonance now appearing at around 450\,nm, due to a very rapid decrease in the gap plasmon index
when the gap width increases (at the groove bottom) from 0.1 to 5\,nm.

\begin{figure}
\centering \includegraphics[width=0.7\columnwidth]{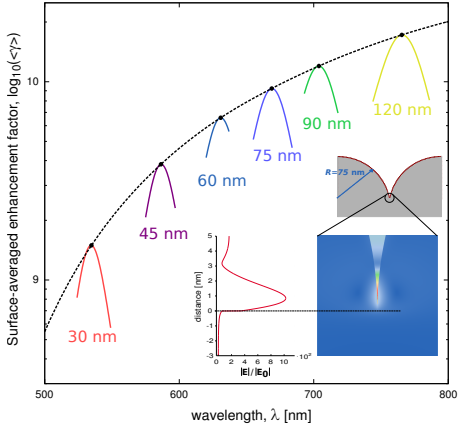}
\caption{Near-resonance plots of the surface-averaged SERS enhancement factor $\left<\gamma\right>$
for arbitrarily well-defined grooves without smoothening ($r=0$) for six cases with $R$ varying
from $30$ to $120$~nm. The inset shows the field-amplitude  distribution $|{\bf E}|/|{\bf E}_{0}|$
for $R=75$\,nm.}\label{plot4}
\end{figure}

With less geometrical smoothening ({\em i.e.} when $r$ is made smaller and smaller) the shortcomings
of the LRA become more severe. The LRA anticipates a monotonously increasing enhancement
factor~\cite{Xiao:2008} and decreasing $r$ also causes a stronger interaction between neighboring
half-cylinders and consequently a redshift~\cite{Garcia-Vidal:1996}. Note that in the interpretation
based on gap surface plasmons~\cite{Sondergaard:2010}, the redshift is simply related to an increase
in the gap plasmon index when the gap width decreases at the groove bottom. In Fig.~\ref{plot3} we
decrease $r$ from 1\,nm down to zero and see how nonlocal effects cause a different trend (indicated
by the dashed line) due to the competing length scales. In particular, for $r\lesssim v_F/\omega$
there is a fundamental saturation of the enhancement factor rather than a monotonous increase and
for our particular choice of the cylinder radius $R$ we see that the $\big<\gamma\big>$ does not
exceed $2\times 10^{9}$.

To explore the ultimate limitations on the SERS in this geometry, Fig.~\ref{plot4} shows results
where we have completely refrained from any geometrical smoothening ($r=0$) and where $v_F/\omega$
is the only length scale that puts fundamental limitations on the field enhancement. As the radius
$R$ of the half-cylinders is increased from 30\,nm to 120\,nm we see a redshift of the peak as also
anticipated in the LRA~\cite{Xiao:2008}. At the same time, the enhancement factor exhibits an
increasing trend where larger cylinders support larger field enhancement by harvesting the incoming
field from larger areas. We emphasize that in all examples the field enhancement remains finite
despite the fact that the crevice is arbitrarily sharp and well defined ($r=0$). For the largest
radius $R$ considered the electromagnetic SERS enhancement factor does not exceed $2\times 10^{10}$.
This illustrates the fundamental limitations imposed by nonlocal response in our specific SERS
configuration.

In conclusion, we have shown that a nonlocal treatment of the plasmonic response leads to new and
possibly fundamental limitations on the electromagnetic SERS enhancement factor, thereby completely changing
the message of the commonly employed local-response approximation of the plasmons. The intrinsic
length scale of the electron gas serves to smear out the field singularity that otherwise would
arise from a local-response treatment and as a consequence the enhancement remains finite even for
geometries with infinitely sharp features. Finally, beyond the linear response fundamental limitations may arise due to nonlinearities~\cite{Ginzburg:2010}.


\end{document}